\documentclass[aps,prl,reprint,superscriptaddress]{revtex4-1}
\usepackage{graphicx}
\usepackage{amsmath}
\usepackage{amssymb}
\usepackage[colorlinks,linkcolor=blue,
            citecolor=blue,
            hyperindex,
            pdfstartview=FitH,
            plainpages=false]
            {hyperref}
\usepackage{soul}

\newcommand {\bscco}{Bi$_2$Sr$_2$CaCu$_2$O$_{8+\delta}$}
\newcommand {\dybscco}{Bi$_2$Sr$_2$Ca$_{0.6}$Dy$_{0.4}$Cu$_2$O$_{8+\delta}$}
\newcommand {\uJcm}{$\mu$J/cm$^{2}$}

\begin{document}
\title{Particle-hole Asymmetry in the Cuprate Pseudogap Measured with Time-Resolved Spectroscopy}

\author{Tristan L. Miller}
\affiliation{Materials Sciences Division, Lawrence Berkeley National Laboratory, Berkeley, California 94720, USA}
\affiliation{Department of Physics, University of California, Berkeley, California 94720, USA}
\author{Wentao Zhang}
\affiliation{Materials Sciences Division, Lawrence Berkeley National Laboratory, Berkeley, California 94720, USA}
\affiliation{Department of Physics and Astronomy, Shanghai Jiao Tong University, Shanghai 200240, China.}
\author{Hiroshi Eisaki}
\affiliation{Electronics and Photonics Research Institute, National Institute of Advanced Industrial Science and Technology, Ibaraki 305-8568, Japan}
\author{Alessandra Lanzara}
\email{alanzara@lbl.gov}
\affiliation{Materials Sciences Division, Lawrence Berkeley National Laboratory, Berkeley, California 94720, USA}
\affiliation{Department of Physics, University of California, Berkeley, California 94720, USA}
\date {\today}


\begin{abstract}

One of the most puzzling features of high-temperature cuprate superconductors is the pseudogap state, which appears above the temperature at which superconductivity is destroyed.  There remain fundamental questions regarding its nature and its relation to superconductivity.  But to address these questions, we must first determine whether the pseudogap and superconducting states share a common property: particle-hole symmetry. We introduce a new technique to test particle-hole symmetry by using laser pulses to manipulate and measure the chemical potential on picosecond time scales. The results strongly suggest that the asymmetry in the density of states is inverted in the pseudogap state, implying a particle-hole asymmetric gap.  Independent of interpretation, these results can test theoretical predictions of the density of states in cuprates.
\end{abstract}

\maketitle

Superconductors have an energy gap representing the binding energy of Cooper pairs, and this gap is particle-hole symmetric in the sense of being centered at the chemical potential.  Cuprate superconductors have another gap above the superconducting critical temperature ($T_c$) known as the pseudogap\cite{Damascelli2003,Norman2005,Hashimoto2014}.  Whether the pseudogap has particle-hole symmetry is at the heart of a dispute about its nature.  Some studies support a particle-hole symmetric pseudogap\cite{Wang2006,Kanigel2007,Kanigel2008,Lee2009,Chatterjee2010,Reber2012,Reber2013}, representing fluctuating superconducting order\cite{Emery1995}.  Other studies claim to demonstrate the particle-hole asymmetry of the pseudogap\cite{Yang2008,Hashimoto2010,He2011,Sakai2013}.  This could be consistent with various theories, including $d$-density wave order theory\cite{Chakravarty2001}, the Yang-Rice-Zhang (YRZ) model of a doped resonant valence bond state\cite{Yang2006,Rice2012, LeBlanc2014}, algebraic charge liquid theory\cite{Qi2010}, Amperean pairing theory\cite{Lee2014}, or an alternate model of fluctuating superconducting order\cite{Grilli2009}.  Several of these theories also purport to explain the nature of superconductivity.

While the superconducting gap by itself has particle-hole symmetry, complete particle-hole symmetry also requires that the electronic density of states is symmetric with respect to the chemical potential. If the density of states is asymmetric, then the chemical potential must adjust with the temperature to conserve the charge. When the density of states is constant with respect to temperature, the chemical potential follows
\begin{equation}
\Delta\mu_\varepsilon \propto -w^2 \frac{D'(E)}{D(E)}\Big|_{E\sim E_F}\label{DOSasymmetry},
\end{equation}
where $w$ is the width of the electronic distribution function (typically proportional to temperature), $D(E)$ is the density of states as a function of energy, and $D'(E)$ is its first derivative, all evaluated near the Fermi energy $E_F$\cite{MarderMu}.  Thus, $\Delta\mu_\varepsilon$ depends on $D'(E)$, which is a direct measure of the asymmetry of the density of states.  A sign change in $D'(E)$ indicates an inversion in the density of states near the Fermi energy.

The temperature dependence of the chemical potential has previously been measured in cuprates using the Kelvin probe technique\cite{Rietveld1990,Rietveld1992,Saito2005,Marel1992}.  However, the Kelvin probe technique only measures the chemical potential relative to the vacuum energy ($\Delta\mu_\mathrm{vac}$), distinct from $\Delta\mu_\varepsilon$ in Eq. (\ref{DOSasymmetry}), which is defined relative to the valence band energy.  Recently, time- and angle-resolved photoemission spectroscopy (TARPES) has been used to measure the change in valence band energy\cite{Rameau2014}, and $\Delta\mu_\varepsilon$\cite{Miller2015a}.  In this technique, $w$ is changed not by adjusting the temperature, but by pumping the sample with a laser pulse.  $w$ is proportional to an effective electronic temperature ($T_e$) which rises and falls on a picosecond time scale.  The function $\Delta\mu_\varepsilon(T_e)$ can be used to characterize the density of states.

Here, we used TARPES to measure $\Delta\mu_\varepsilon$ in high temperature superconductors \bscco{} (Bi2212) and \dybscco{} (Dy-Bi2212) over a large doping range both inside and outside the pseudogap region.  Our results on optimally and overdoped samples are consistent with the density of states in these materials.  However, in an underdoped sample, the function $\Delta\mu_\varepsilon(T_e)$ is inverted, suggesting a sign change in $D'(E)$.  This sign change cannot be caused by a particle-hole symmetric gap.  We instead propose that a particle-hole asymmetric pseudogap introduces an anomaly in the density of states just above the Fermi energy.  We use the YRZ model\cite{Yang2006, Rice2012,LeBlanc2014} to illustrate this scenario, and discuss other possible scenarios.

Single crystals of Bi2212 and Dy-Bi2212 were cleaved \em{in situ}\em{} in an ultrahigh vacuum.  Samples were pumped with 1.48 eV laser pulses and electrons were photoemitted with 5.93 eV laser pulses, with a time resolution of $\sim$ 300 fs and energy resolution of $\sim$ 22 meV.  Measurements were taken in cycles so that the long-term drift in the chemical potential could be corrected.  All error bars are estimated from the variance between cycles.  The experimental methods are identical to those in Ref. \cite{Miller2015a}.  

\begin{figure*}\centering\includegraphics[width=7in]{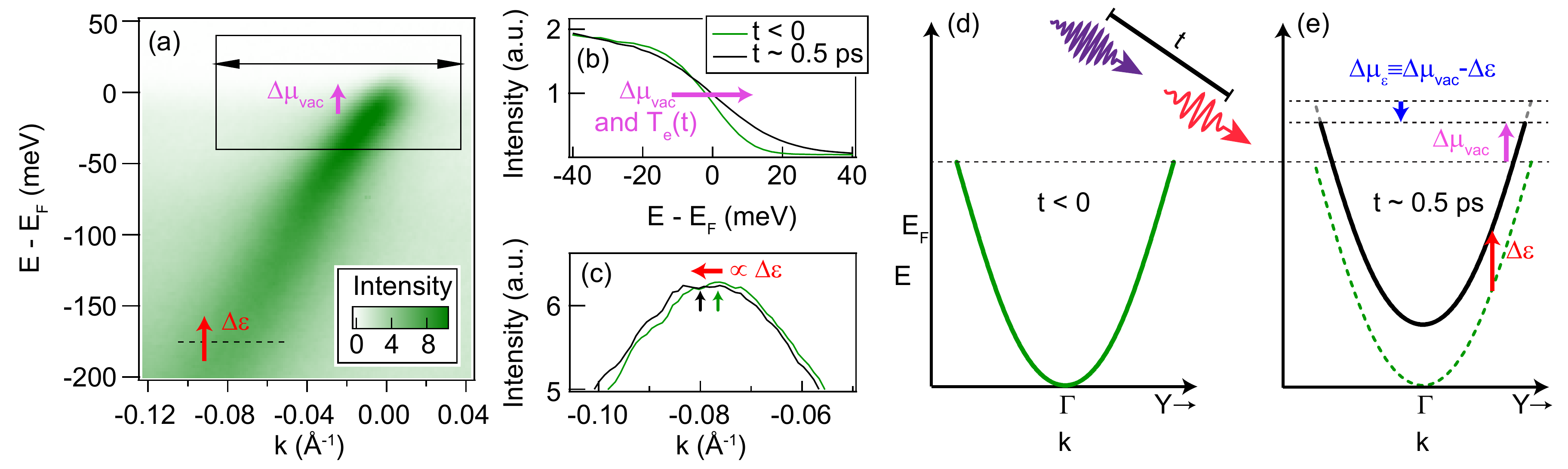}
\caption{(a) The momentum-energy map of ARPES intensity on overdoped Bi2212 along the $\Gamma-Y$ momentum direction.  (b) The ARPES intensity integrated along the momentum range indicated by the double arrows in (a) both before pumping and after pumping.  $\Delta\mu_\mathrm{vac}$ and $T_e$ are obtained by fitting the edge.  (c) The intensity along the dotted line indicated in (a) both before and after pumping.  $\Delta\varepsilon$ is obtained from the shift in peak position.  (d,e) An illustration (not to scale) of the valence band dispersion along $\Gamma-Y$ and its response to pumping.}
\label{Fig1}
\end{figure*}

Figure \ref{Fig1} establishes the procedure for finding $\Delta\mu_\varepsilon$ with TARPES, starting with an overdoped (OD) Bi2212 sample ($T_c \sim{}$ 70 K), where the pseudogap is small or absent\cite{Vishik2012}.  Panel (a) shows a momentum-energy map of photoemission intensity measured along the $\Gamma-Y$ momentum direction at a low temperature ($T$=30 K).   The chemical potential is a geometry-independent property, but the $\Gamma-Y$ geometry is convenient since here the superconducting gap vanishes.   Energies are shown relative to the estimated Fermi energy ($E_F$), which is the chemical potential at zero temperature.  The ARPES intensity taken before pumping ($t <$ 0) and after pumping ($t \sim$ 0.5 ps) is integrated along the momentum region shown by the horizontal black arrow the in panel (a).  The results, shown in Fig. \ref{Fig1}(b), are fit with a Fermi-Dirac distribution\cite{Miller2015a}.   $\mu_\mathrm{vac}$ and $T_e$ are both parameters of the fit, and $\Delta\mu_\mathrm{vac}$ is defined as the change in $\mu_\mathrm{vac}$ upon pumping.  Next, we extract the valence band dispersion using standard methods\cite{LaShell2000}.  At each energy, momentum distribution curves (MDCs) are extracted [see Fig. \ref{Fig1}(c)] and fit to Lorentzian curves, and $\Delta\varepsilon$ is determined from the pump-induced shift in the MDC peak positions in the 150--200 meV range. $\Delta\varepsilon$  represents the change in band dispersion upon pumping [see panel (e)] and can be understood as the photodoping effect investigated elsewhere\cite{Rameau2014}.

The shift in chemical potential relative to the valence band ($\Delta\mu_\varepsilon$) can finally be extracted from the difference $\Delta\mu_\mathrm{vac} - \Delta\varepsilon$, as illustrated in Figs. \ref{Fig1}(d) and \ref{Fig1}(e).  Panel (d) shows an illustration of the full valence band dispersion along $\Gamma-Y$, and panel (e) shows the same dispersion at about 0.5 ps after pumping.  The valence band shifts by $\Delta\varepsilon$, while the chemical potential shifts by $\Delta\mu_\mathrm{vac}$. The difference $\Delta\mu_\varepsilon$ is a quantity of distinct origin\cite{Miller2016SI}, which provides information about the density of states.

\begin{figure}\centering\includegraphics[width=3.4in]{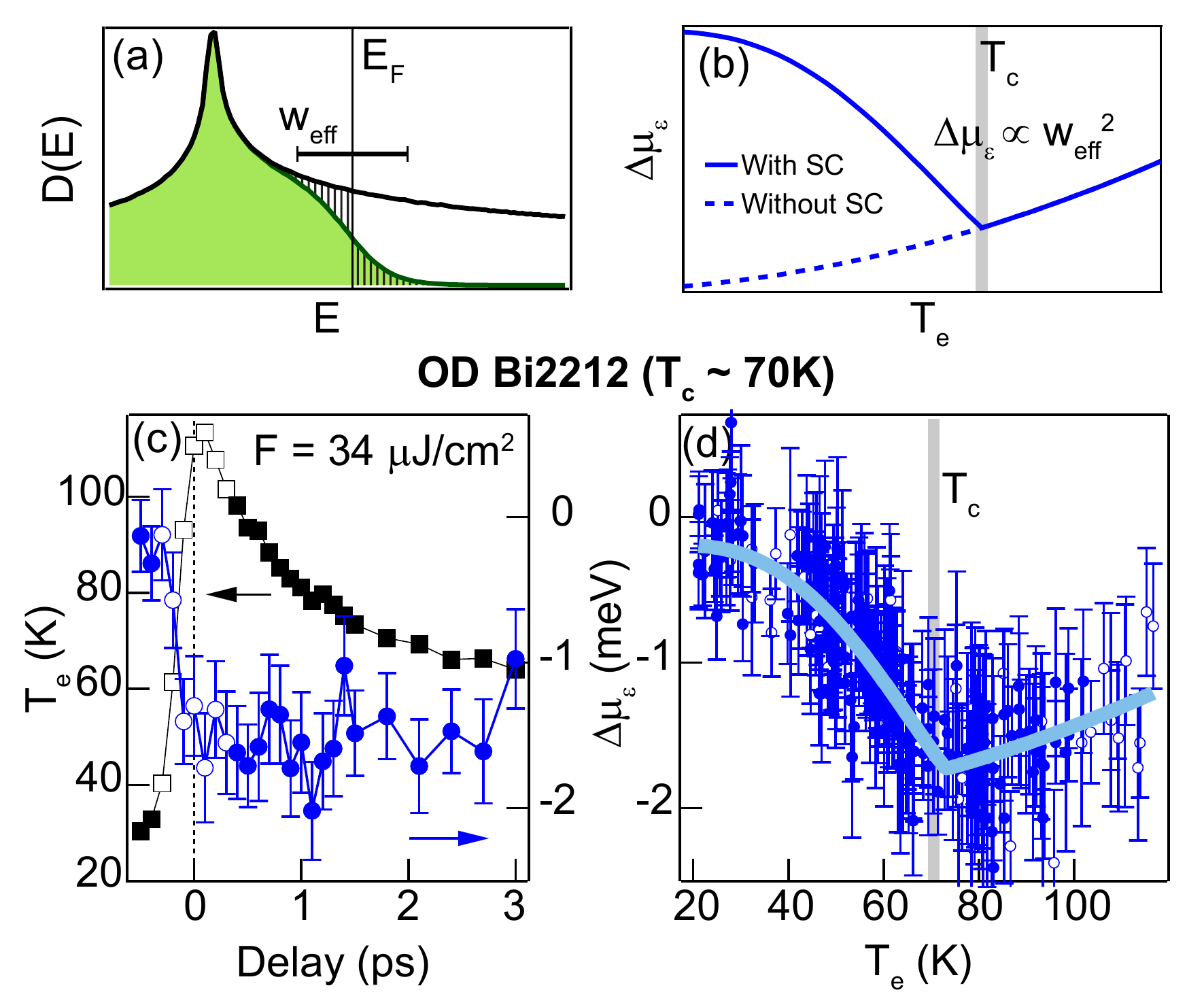}
\caption{(a) The density of states in a Bi2212 sample in the normal state. (b) The expected function $\Delta\mu_\varepsilon(T_e)$ both with and without the superconducting (SC) gap.  (c) $\Delta\mu_\varepsilon$ and $T_e$ as a function of delay time $t$ after pumping an overdoped Bi2212 sample with a 34 \uJcm{} pulse. (d) $\Delta\mu_\varepsilon$ is shown as a function of $T_e$, eliminating $t$ as a variable.  Data points are collected from delay times up to 10 ps, pump fluences between 3 and 34 \uJcm{}, and initial temperatures between 25 and 80 K.  Empty circles denote the data collected between a --0.3 and 0.3 ps delay.
}
\label{Fig2}
\end{figure}

To better understand $\Delta\mu_\varepsilon$ we first explain the prediction from the basic considerations of the density of states in Bi2212.  Figure \ref{Fig2}(a) shows the density of states in Bi2212 from a phenomenological tight-binding model of the valence band\cite{Norman1995}.  Because of the saddle points at the $M$ points of the Brillouin zone, there is a van Hove singularity, seen as a peak in the density of states below the Fermi energy.  The number of electrons at a given energy is
\begin{equation}
 N(E) = D(E) f(E-\mu,T_e)  \label{Electron population},
\end{equation}
where $f(E-\mu,T_e)$ is the Fermi-Dirac distribution at $T_e$, and all energies are measured relative to the valence band energy. The total number of electrons $N$ is equal to the sum of $N(E)$ over all energies [the area of the green shaded region in Fig. \ref{Fig2}(a)].  By the conservation of charge, $N$ must be constant as the temperature changes, and therefore $\mu_\varepsilon$ must shift until the two striped regions in Fig. \ref{Fig2}(a) are of equal area.  This leads to a derivation of Eq. (\ref{DOSasymmetry}).  The order of magnitude of $\mu_\varepsilon$ can also be estimated as $(k_B T_c)^2 \frac{D'(E)}{D(E)}\Big|_{E\sim E_F}$, which is about 0.5 meV for Bi2212 near optimal doping.

This tight-binding model does not include the superconducting gap, which causes $D(E)$ to change with the temperature, rendering Eq. (\ref{DOSasymmetry}) inapplicable.  However, as argued in Ref. \cite{Miller2015a}, it is possible to change variables such that we consider the quasiparticle energy with the pairing interaction turned off.  Under this change of variables, $f(E-\mu,T_e)$ is no longer a Fermi-Dirac distribution but now has an effective width $w_{\mathrm{eff}}$ proportional to the size of the superconducting gap ($\Delta_{\mathrm{SC}}$) even at zero temperature.  This enhanced width does not represent thermal excitations but rather the Bogoliubov mixing of electronlike and holelike excitations.   Pumping is known to suppress the superconducting gap\cite{Smallwood2014,Smallwood2016}; thus, as we approach $T_e = T_c$ the value of $w_{\mathrm{eff}}$ may actually decrease with the temperature, if $\Delta_{\mathrm{SC}} >> k_BT_c$.  Based on these considerations Fig. \ref{Fig2}(b) shows, qualitatively, the expected $\Delta\mu_\varepsilon(T_e)$.

Figures \ref{Fig2}(c) and \ref{Fig2}(d) compare this simple model to our experimental results on OD Bi2212.  Panel (c) shows $\Delta\mu_\varepsilon$ and $T_e$ as functions of delay time $t$ after pumping a sample at 30 K with a 34 \uJcm{} pulse.  In panel (d) we plot $\Delta\mu_\varepsilon$ directly as a function of $T_e$, showing good agreement with the theoretical model [shown in panel (b)].  Data were collected for $t$ up to 10 ps, combining five different experiments with pump fluences ranging from 3 to 34 \uJcm{} and initial temperatures ranging from 25 to 80 K.  Each experiment was shifted by a constant energy to account for the initial chemical potential at different initial temperatures.  These data demonstrate the nontrivial fact that the function $\Delta\mu_\varepsilon(T_e)$ exists independently of changes in experimental conditions.  In particular, we show that the data taken between -0.3 and 0.3 ps (open circles) fall along the same curve, reinforcing the notion $\Delta\mu_\varepsilon(T_e)$ is an equilibrium property of the system rather than a dynamical property.  This is consistent with previous studies on Bi2212 showing that the electrons thermalize within 100 fs after pumping\cite{Perfetti2007,Graf2011}.


\begin{figure*}\centering\includegraphics[width=7in]{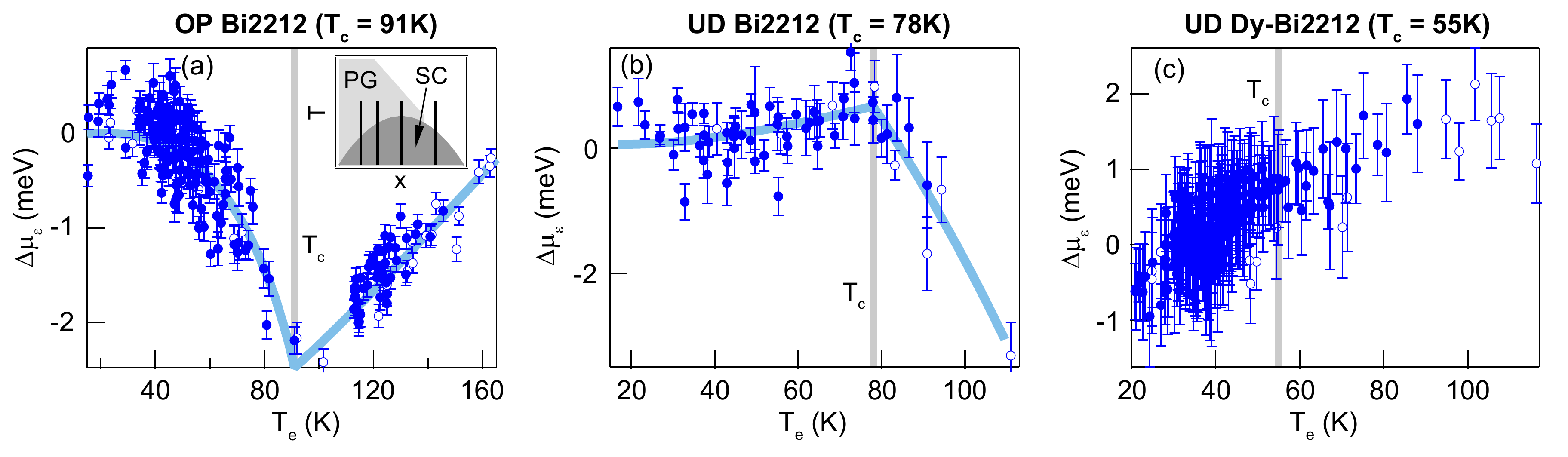}
\caption{ The function $\Delta\mu_\varepsilon(T_e)$ is plotted for OP Bi2212 with $T_c$ = 91 K (a), a slightly UD Bi2212 with $T_c$ = 78 K (b), and a very UD Dy-Bi2212 with $T_c$ = 55 K (c).  Data are collected over many delay times, pump fluences, and initial temperatures.  The inset of (a) shows the cuprate phase diagram, including the superconducting (SC) and pseudogap (PG) phases, with black lines indicating the four samples in this study.  Empty circles denote the data taken between a --0.3 and 0.3 ps delay, and gray bars indicate $T_c$.  (a) uses data from Ref. \cite{Miller2015a}.
}
\label{Fig3}
\end{figure*}

Since the density of states of Bi2212 will be affected by the pseudogap and its size relative to other energy scales, we extended our measurement to more and more underdoped samples, which are known to have larger and larger pseudogaps\cite{Zhao2013,Vishik2012}.  Figure \ref{Fig3} shows the results on optimally doped (OP) Bi2212 ($T_c$ = 91 K) slightly underdoped (UD) Bi2212 ($T_c$ = 78 K), and very UD Dy-Bi2212 ($T_c$ = 55 K).  The trends in $\Delta\mu_\varepsilon(T_e)$ are qualitatively distinct in each case.  This suggests that the relevant property of the material is not just the size of the pseudogap, but its relative size compared to another energy scale.  In this picture, the three distinct regimes arise when the pseudogap is comparatively large, small, or just the right size.

Examining the trends in more detail, we find that in OP Bi2212 [Fig. \ref{Fig3}(a)] (small pseudogap regime), $\Delta\mu_\varepsilon$ decreases with temperature up to $T_c$ and increases beyond that.  This is similar to the trends in OD Bi2212 [Fig. \ref{Fig2}(d)], and it aligns with our expectations in the absence of the pseudogap.  In  slightly UD Bi2212 [Fig. \ref{Fig3}(b)] (midsized pseudogap regime), $\Delta\mu_\varepsilon$ follows an inverted trend, increasing with temperature up to $T_c$, and decreasing with temperature beyond that.  Following Eq. (\ref{DOSasymmetry}), this suggests that the sign of $D'(E)$ is changed, which implies that the pseudogap is particle-hole asymmetric.   Finally, in heavily UD Dy-Bi2212 [Fig. \ref{Fig3}(c)] (large pseudogap regime), $\Delta\mu_\varepsilon$ increases with temperature both below and above $T_c$.  The lack of change across $T_c$ suggests that the effect of the superconducting gap is washed out by a sufficiently large pseudogap.  The sign in the trend suggests that $D'(E)$ changes sign only when the pseudogap is the right size, and not when it is either too large or too small.

\begin{figure}\centering\includegraphics[width=3.4in]{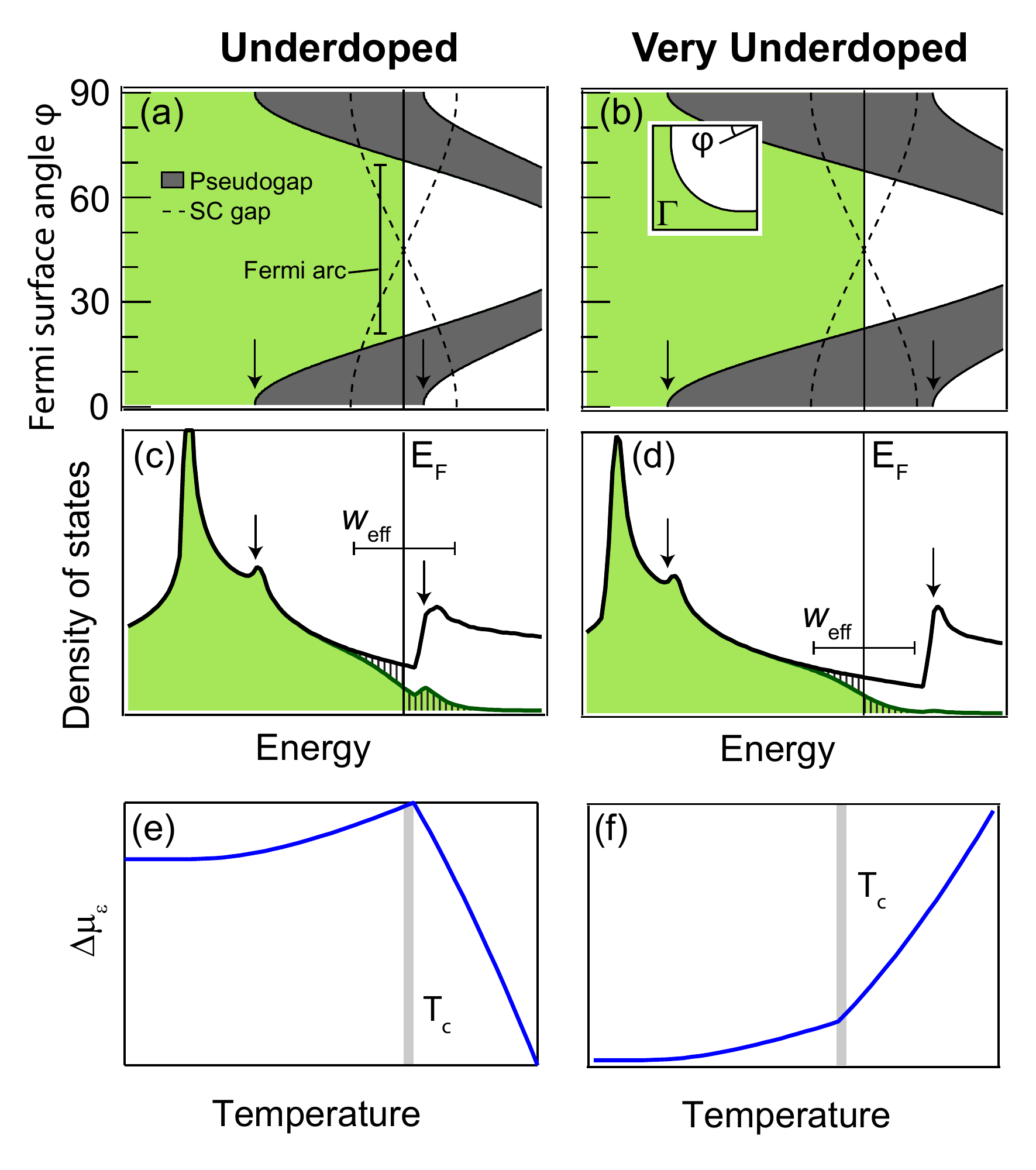}
\caption{ An illustration of how the YRZ model is potentially consistent with results on the chemical potential in underdoped (a),(c),(e) and very underdoped (b),(d),(f) Bi2212.  (a),(b) A diagram of the pseudogap, parametrized by the Fermi surface angle $\varphi$ shown in the quarter Brillouin zone in the inset of (b).   The very underdoped sample has a larger pseudogap.  (c),(d) The calculated density of states, with the green shaded region indicating occupied states.  Arrows indicate the edges of the antinodal pseudogap.  (e),(f) Qualitative predictions of the function $\mu_\varepsilon(T_e)$.  Gray bars indicate $T_c$.}
\label{Fig4}
\end{figure}

We note that, although a sign change in $D'(E)$ points to a particle-hole asymmetric gap, not every particle-hole asymmetric model would necessarily produce the behavior observed here.  In Fig. 4 we compare our experimental findings to a relatively simple particle-hole asymmetric pseudogap model, the YRZ model\cite{Yang2006,Rice2012,LeBlanc2014}.  Figures \ref{Fig4}(a) and \ref{Fig4}(b) show the pseudogap in two underdoped samples as a function of the Fermi surface angle ($\varphi$).  Near the node ($\varphi \sim 45^\circ$), the pseudogap is entirely above the Fermi energy, resulting in the well-known Fermi arc\cite{Loeser1996,Marshall1996}.  Near the antinodes ($\varphi = 0$ and $90^\circ$), the pseudogap is centered below the Fermi energy.  For the slightly underdoped sample [panel (a)], the antinodal upper pseudogap edge lies just above the Fermi energy, but for the very underdoped sample [panel (b)], the pseudogap is larger, such that the antinodal upper pseudogap edge is further from the Fermi energy.  Note that previous studies have little to say about the precise location of the upper antinodal pseudogap edge, since ARPES, the primary means of measuring and resolving the momentum dependence of the pseudogap, can only directly measure the lower pseudogap edge.

Figures \ref{Fig4}(c) and \ref{Fig4}(d) show density of states calculations for each pseudogap size.  There are clear anomalies in the density of states at the upper and lower antinodal gap edges (black arrows).  For the slightly underdoped sample [Fig. \ref{Fig4}(c)], the anomaly just above $E_F$ causes $D'(E)$ to have a positive sign near $E_F$, which leads to an inversion of $\Delta\mu_\varepsilon(T_e)$ [Fig. \ref{Fig4}(e)].  This is similar to the inversion seen in UD Bi2212 [Fig. \ref{Fig3}(b)].  However, for the very underdoped sample [Fig. \ref{Fig4}(d)], the anomaly is too far away from $E_F$ to impact $D'(E)$ near $E_F$.  Furthermore, the large antinodal pseudogap restricts the impact of the superconducting gap to the Fermi arc, and at these momenta the superconducting gap is smaller [see Fig. \ref{Fig4}(b)].  This reduces the effect of superconductivity on $w_{\mathrm{eff}}$, resulting in a relatively weak feature at $T_c$ in $\Delta\mu_\varepsilon(T_e)$ [Fig. \ref{Fig4}(f)].  This is similar to our results on underdoped Dy-Bi2212 [Fig. \ref{Fig3}(c)].

The results of this study show that the slope in density of states near the Fermi energy changes sign in Bi2212 in the presence of a pseudogap.  This implies that the pseudogap has particle-hole asymmetry, unlike the symmetry of the superconducting gap.  The sign change occurs both above and below $T_c$, supporting previous work showing that the pseudogap coexists with superconductivity\cite{Kondo2007,Ma2008,Khasanov2008,Kondo2009,Vishik2012,Hashimoto2015}.  Using calculations of the YRZ model\cite{Yang2006,Rice2012, LeBlanc2014}, we have shown that these results are well described by an anomaly in the density of states just above the Fermi energy, created by the antinodal upper pseudogap edge.  Our results on very underdoped Dy-Bi2212 are described by an anomaly that appears further away from the Fermi energy.  These results may also be consistent with other particle-hole asymmetric pseudogap models (e.g. the Amperean pairing model\cite{Lee2014}), but this must be verified by calculations of the density of states.

We note that our conclusion of particle-hole asymmetry relies on several assumptions.  First, we used Eq. (\ref{DOSasymmetry}) rather than the more general Eq. (\ref{Electron population}), ignoring the effects of a temperature-dependent density of states. There is a remote possibility that the pseudogapped density of states changes with temperature, in precisely the right way, to give only the appearance of an asymmetric pseudogap.  Second, underdoped cuprates are also known to have a charge density wave state\cite{Comin2014} and a $d$-symmetry form factor density wave state\cite{Hamidian2015}, and it is unknown whether these states are related to the pseudogap.  Third, while we have shown qualitative consistency with the YRZ model, further work is needed to show quantitative consistency with this or other pseudogap models.

An exciting implication of our results is that the upper edge of the antinodal pseudogap can be directly accessed by ARPES because it falls within the effective quasiparticle distribution width $w_{\mathrm{eff}}$.  Note that $w_{\mathrm{eff}}$ is dominated by Bogoliubov mixing rather than thermal excitations, and so, we predict that excitations above the Fermi energy will be reflected below the Fermi energy, seen as a small peak or shoulder in the ARPES spectra\cite{LeBlanc2011}.  This may have already been observed\cite{He2011}.  Study of this feature may provide more detailed momentum-dependent information on the pseudogap.

\begin{acknowledgments}
We thank J. P. F. LeBlanc and J. W. Orenstein for helpful feedback. This work was supported by the U.S. Department of Energy, Office of Science, Basic Energy Sciences, Materials Sciences and Engineering Division under Contract No. DE-AC02-05-CH11231 within the Ultrafast Materials Science Program (KC2203).
\end{acknowledgments}

\bibliographystyle{apsrev4-1}

\end{document}